\documentclass[ reprint,nofootinbib,amsmath,amssymb,onecolumn,aps]{revtex4-2}
\usepackage{appendix}
\usepackage{graphicx}
\usepackage{dcolumn}
\usepackage[colorlinks,linkcolor=magenta,anchorcolor=cyan,citecolor=blue]{hyperref}
\usepackage{physics}
\usepackage{bm}

\def\be{\begin{equation}}
\def\ee{\end{equation}}
\def\ba{\begin{eqnarray}}
\def\ea{\end{eqnarray}}

              \def\.{\cdot}

\begin{document}
\title{The identification between the bulk and boundary conserved quantities}
\author{Gerui Chen$^{1,2}$}
\email{t20152277@csuft.edu.cn}
\author{Zien Gao$^{2,3}$}
\email{zegao@mail.bnu.edu.cn}
\author{Xin Lan$^{2,3}$}
\email{xinlan@mail.bnu.edu.cn}
\author{Jie-qiang Wu$^{4,5}$}
\email{jieqiangwu@itp.ac.cn}
\author{Hongbao Zhang$^{2,3}$}
\email{hongbaozhang@bnu.edu.cn}
\affiliation{$^1$ College of Electronic Information and Physics, Central South University of Forestry and Technology, Changsha 410004, China\\
$^2$School of Physics and Astronomy, Beijing Normal University, Beijing 100875, China\\
$^3$ Key Laboratory of Multiscale Spin Physics, Ministry of Education, Beijing Normal University, Beijing 100875, China\\
$^4$ CAS Key Laboratory of Theoretical Physics, Institute of Theoretical Physics, Chinese Academy of
Sciences, Beijing 100190, China\\
$^5$ School of Physical Sciences, University of Chinese Academy of Sciences, Beijing 100049, China}

\date{\today}

\begin{abstract}
By using Wald formalism, we show that the identification between the bulk and boundary conserved quantities induced by the perturbation of generic non-electromagnetic matter field holds not only on top of the asymptotically flat stationary spacetimes but also on top of the asymptotically AdS stationary ones.  We further show that such an identification reduces to the familiar form for the test point particle by viewing it as the limiting case of general matter.

\end{abstract}
\maketitle
\section{Introduction}
In physics without gravity included, for instance in special relativity, the conserved quantities associated with the spacetime Poincare symmetry can be defined as the integral on a spacelike Cauchy surface of the conserved current obtained from the energy momentum tensor contracted with the Killing field.  However, in the presence of gravity, it is well known that the sensible conserved quantities can only be defined at infinity, instead associated with the asymptotic spacetime symmetry. Although they appear to be obviously different from each other in the sense that the former is given by an integral in the bulk while the latter is evaluated on the boundary, the identification of these two kinds of conserved quantities is used extensively for those scenarios in which the matter fields are treated as the perturbed source to the background spacetime. But such an identification is generically assumed a priori without derivation.
To the best of our knowledge, \cite{GW} is supposed to be the first to suggest a general proof of such an identification for the perturbed matter fields on top of arbitrary asymptotically flat stationary spacetimes by working with Wald formalism. On the other hand, \cite{RC} provides a specific proof for the test ring in Banados-Teitelboim-Zanelli (BTZ) black hole through solving the perturbed Einstein equation in an explicit way. Although the original goal in \cite{RC} is to seek a proof for the test point particle, the reason why they instead work with the test ring is to keep the axisymmetry of the system such that the perturbed Einstein equation can be readily solved.
Later on, this strategy is further adopted in the similar proof for the test shell in other black hole backgrounds\cite{RS,RSD}.

The purpose of this paper is twofold. First, by following the strategy suggested in \cite{GW}, we shall make use of Wald formalism to prove that the aforementioned identification for the perturbed matter fields holds not only on top of the asymptotically flat stationary spacetimes but also on top of the asymptotically Anti-de Sitter (AdS) stationary ones. It is noteworthy that we are not required to solve the perturbed Einstein equation explicitly at all such that the resulting proof is not restricted to the specific matter perturbation considered in \cite{RC,RS,RSD}, but applicable to generic matter perturbation. In light of this, we further derive the corresponding result for the point particle by viewing it as the limiting case of general matter.



The structure of this paper is organized as follows. After a brief review of the fundamental identity in Wald formalism, in the subsequent section we shall apply it to achieve an elegant proof of the above identification between the bulk and boundary conserved quantities for the generic matter perturbation on top of the stationary spacetime which is either asymptotically flat or asymptotically AdS. In Section \ref{pp}, after a brief presentation of the action of a point particle and its dynamics, we develop a relationship between the field theory and worldline description of the point particle by virtue of the diffeomorphism invariance of the action, whereby we further derive the aforementioned identification for the point particle. We shall conclude our paper with some discussions in the final section.

We will follow the notation and conventions of \cite{GR}. In particular, early Latin letters denote the abstract indices, and Greek letters denote the specific indices. In addition, we shall use the boldface letters to denote differential forms with the tensor indices suppressed.

\section{The bulk and boundary identification for general matter}
The dynamics of gravitational and electromagnetic field is assumed to be governed by the Einstein-Maxwell Lagrangian form as follows
\begin{equation}
    \mathbf{L}=\frac{1}{16\pi}\bm{\epsilon}(R-2\Lambda-F_{ab}F^{ab}),
\end{equation}
where $\bm{\epsilon}$ is the spacetime volume compatible to the metric, $\Lambda$ is the cosmological constant, and $\mathbf{F}=d\mathbf{A}$ is the field strength of the electromagnetic field potential $\mathbf{A}$.
Its variation can be written as
\begin{equation}
    \delta \mathbf{L}=\bm{\epsilon}E(\phi)\delta\phi+d\mathbf{\Theta}(\phi;\delta\phi),
\end{equation}
where $g_{ab}$ and $A_a$ have been denoted collectively by $\phi$ and $\delta$ will be understood as the exterior differential in the configuration space below. Specifically, we have
\begin{equation}
   - \frac{1}{2}T^{ab}\equiv E^{ab}=-\frac{1}{16\pi}(G^{ab}+\Lambda g^{ab})+\frac{1}{8\pi}(F^{ac}F^b{}_c-\frac{1}{4}g^{ab}F^{cd}F_{cd}), \quad -j^a\equiv E^a=\frac{1}{4\pi}\nabla_b F^{ba},
\end{equation}
whereby $T^{ab}$ and $j^a$ correspond virtually to the non-electromagnetic energy momentum tensor and electric current, respectively. In addition, we have $\mathbf{\Theta}=\mathbf{\Theta}^{GR}+\mathbf{\Theta}^{EM}$ with
\begin{equation}
    \Theta^{GR}_{bcd}=\frac{1}{16\pi}\epsilon_{abcd}g^{ae}g^{fg}(\nabla_g\delta g_{ef}-\nabla_e\delta g_{fg}),\quad \Theta^{EM}_{bcd}=-\frac{1}{4\pi}\epsilon_{abcd}F^{ae}\delta A_e.
\end{equation}

The Noether current associated with the gauge transformation $\delta_\chi A_a=\nabla_a\chi$ is defined as
\begin{equation}
    \mathbf{J}_\chi=X_\chi\cdot\mathbf{\Theta},
\end{equation}
where $X_\chi\equiv\int d^4x\sqrt{-g}\nabla_a\chi\frac{\delta}{\delta A_a}$ is understood as the vector field in the configuration space and the dot denotes the contraction with the first index of the involved form. Then one can show that
\begin{equation}
    d\mathbf{J}_\chi=d\mathbf{C}_\chi-\bm{\epsilon}\nabla_aj^a\chi
\end{equation}
with $(C_\chi)_{bcd}=\epsilon_{abcd}j^a\chi$. The arbitrariness of $\chi$ implies that
\begin{equation}\label{another}
   \nabla_aj^a=0,\quad  \mathbf{J}_\chi=d\mathbf{Q}_\chi+\mathbf{C}_\chi
\end{equation}
with $Q_{\chi cd}=-\frac{1}{8\pi}\epsilon_{abcd}F^{ab}\chi$ the Noether charge associated with the gauge transformation induced by $\chi$.

On the other hand, the Noether current associated with the infinitesimal diffeomorphism generator $\xi$ is defined as
\begin{equation}\label{ncdef}
    \mathbf{J}_\xi\equiv X_\xi\cdot\mathbf{\Theta}-\xi\cdot\mathbf{L}
\end{equation}
with $X_\xi\equiv\int d^4x\sqrt{-g}\mathcal{L}_\xi\phi\frac{\delta}{\delta\phi(x)}$, whereby one can show that
\begin{equation}
    d\mathbf{J}_\xi=-\bm{\epsilon}E(\phi)\mathcal{L}_\xi\phi=d\mathbf{C}_\xi+\bm{\epsilon}(-\nabla_aT^{ab}+j_aF^{ba}-\nabla_aj^aA^b)\xi_b,
\end{equation}
where  $\mathbf{C}_\xi$ is given by
\begin{equation}
    (C_\xi)_{bcd}=\epsilon_{abcd}(T^{ae}+j^aA^e)\xi_e.
    \end{equation}
By the same token, it follows from the arbitrariness of $\xi$ that
\begin{equation}\label{divergence}
    \nabla_aT^{ab}+j_aF^{ab}=0,
\end{equation}
and
\begin{equation}\label{ncder}
    \mathbf{J}_\xi=d\mathbf{Q}_\xi+\mathbf{C}_\xi
\end{equation}
with $\mathbf{Q}_\xi=\mathbf{Q}^{GR}_\xi+\mathbf{Q}^{EM}_\xi$ the corresponding Noether charge, given by
\begin{equation}
    Q^{GR}_{ab}=-\frac{1}{16\pi}\epsilon_{abcd}\nabla^c\xi^d,\quad Q^{EM}_{ab}=-\frac{1}{8\pi}\epsilon_{abcd}F^{cd}A_e\xi^e.
\end{equation}
Acting on both Eq. (\ref{ncdef}) and Eq. (\ref{ncder}) with $\delta$, we obtain
\begin{equation}\label{fi}
    d(\delta\mathbf{Q}_\xi-\xi\cdot\mathbf{\Theta})=-X_\xi\cdot\delta \mathbf{\Theta} -\xi\cdot \bm{\epsilon}E(\phi)\delta \phi-\delta\mathbf{C}_\xi,
\end{equation}
which is sometimes called the fundamental identity in Wald formalism
\cite{GW,Wald,IW,SW}.

\begin{figure}
\centering
\includegraphics[width=0.5\textwidth]{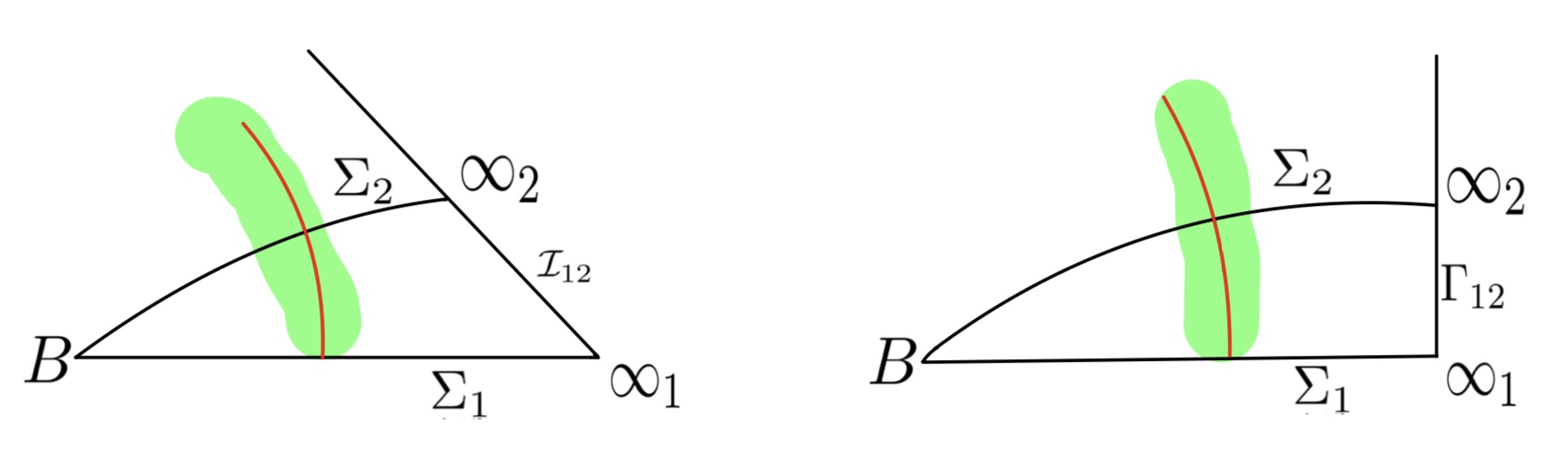}
\caption{The background is perturbed by the non-electromagnetic matter, where the green shaded region and red line represent the distribution for the general matter and the trajectory of the point particle, respectively. The left plot is for the asymptotically flat spacetime, where the hypersurface can end on either the spatial infinity as $\Sigma_1$ or on the null infinity as $\Sigma_2$. The right plot is for the asymptotically AdS spacetime, where the hypersurface can end anywhere on the AdS boundary. The inner boundary $B$ of the hypersurface, if exists, is assumed not to be perturbed by the presence of the matter.   }\label{p}
\end{figure}

As illustrated in Fig.\ref{p}, in what follows we shall consider the perturbation of non-electromagnetic matter on top of the background fields, which are assumed to be stationary and satisfy the Einstein-Maxwell equations $E(\phi)=0$. The background spacetime can be either asymptotically flat or asymptotically AdS. In addition, we also assume that the non-electromagnetic matter is distributed inside of the bulk spacetime, namely absent from the infinity. As such, we choose any Killing vector field as $\xi$ such that $X_\xi=0$\footnote{Here we also require $\mathcal{L}_\xi A_a=0$.}, and perform an integral of Eq. (\ref{fi}) over a hypersurface $\Sigma$. For an asymptotically flat spacetime, the outer boundary of $\Sigma$ can be located at either spatial infinity or future null infinity. In addition, if the background spacetime is a black hole, the inner boundary of $\Sigma$ is chosen to be at a sufficiently early cross section of the black hole event horizon like the bifurcation surface for the non-extremal black hole such that no perturbation is induced over there by the perturbed non-electromagnetic matter\cite{GW,SW}. Otherwise, there is no inner boundary for $\Sigma$. As a result, the above integral yields
\begin{equation}
  \int_\infty ( \delta \mathbf{Q}_\xi-\xi\cdot\mathbf{\Theta})=\int_\Sigma \epsilon_{abcd}(\delta T^{ae}+\delta j^aA^e)\xi_e.
\end{equation}
First, we like to work with the gauge in which $A_a\rightarrow 0$ at infinity, then there is no contribution to the left handed side of the above equation from the electromagnetic part for both asymptotically flat and asymptotically AdS spacetimes\cite{SW,HIM}. Second, not only can such a gravitational contribution be written as $\delta H_\xi$ with $H_\xi$ the conserved quantity associated with $\xi$ at both spatial infinity and future null infinity for asymptotically flat spacetimes with $H_\xi=0$ for Minkowski spacetime\cite{IW,WZ}, but also at spatial infinity for asymptotically AdS spacetimes with $H_\xi=0$ for the pure AdS\cite{HIM}. Thus we end up with the following identification between the bulk and boundary conserved quantities induced by the perturbed non-electromagnetic matter\footnote{The ambitious readers might expect that the bulk conserved quantities could also be expressed as $\int_\Sigma\epsilon_{abcd}(\delta T^{ae}+\delta T^{ae}_{EM})\xi_e$ with $T^{ab}_{EM}$ the energy momentum tensor for the electromagnetic field. However, this is impossible unless there is no background electromagnetic field, in which $\delta T^{ab}_{EM}=0$ indeed. The reason comes from the fact that the current $(\delta T^{ae}+\delta T^{ae}_{EM})\xi_e$ is not conserved because $\nabla_a \delta T^{ab}+\nabla_a \delta T^{ab}_{EM}=-\delta \nabla_a)T^{ab}_{EM}\neq 0$ generically in the presence of the background electromagnetic field.}
    \begin{equation}\label{final}
  \delta H_\xi=\int_\Sigma \epsilon_{abcd}(\delta T^{ae}+\delta j^aA^e)\xi_e,
\end{equation}
which was previously established in \cite{GW} only for asymptotically flat stationary background spacetimes and is now generalized to asymptotically AdS ones.
  It is noteworthy that both sides of the above equation are independent of the choice of $\Sigma$ as it should be the case. Such an independence of the left handed side can be seen readily by evaluating the fundamental identity (\ref{fi}) on the portion of the infinity sandwiched by the outer boundaries of the two hypersurfaces. On the other hand, we have
\begin{equation}
\nabla_a[(T^{ae}+j^aA^e)\xi_e]=\nabla_aT^{ae}\xi_e+j^a\nabla_a(A_e\xi^e)=\nabla_aT^{ae}\xi_e+j^aF_{ae}\xi^e=0,
\end{equation}
where $\nabla_{(a}\xi_{b)}=0$ and the first equation of Eq. (\ref{another}) are used in the first step, $0=\mathcal{L}_\xi\mathbf{A}=\xi\cdot d\mathbf{A}+d(\xi\cdot\mathbf{A})$ is used in the second step, and Eq. (\ref{divergence}) is used in the last step. This further implies the independence of the right handed side on the choice of $\Sigma$.

\section{The bulk and boundary identification for point particle}\label{pp}
Let us start with the action of a point particle with mass $m$ and charge $q$ in electromagnetic and  gravitational fields as follows
\begin{equation}\label{pa}
    S_p=\int \mathbf{L}_p=-m\int d\tau \sqrt{-g_{\mu\nu}\frac{dx^\mu}{d\tau}\frac{dx^\nu}{d\tau}}+q\int d\tau A_\mu\frac{dx^\mu}{d\tau},
\end{equation}
where $\mathbf{L}_p$ represents the Lagrangian one form of the point particle. Note that the Lagrangian one form is independent of the choice of the time parameter $\tau$. So for simplicity but without loss of generality, we shall choose the proper time as the time parameter for a given worldline, namely $g_{\mu\nu}U^\mu U^\nu=-1$ with the four velocity $U^\mu=\frac{dx^\mu}{d\tau}$. Now the variation of such a worldline gives
\begin{equation}
    \delta \mathbf{L}_p=d\tau E_\nu\delta x^\nu+d\theta,
\end{equation}
where
\begin{equation}
    E_\nu=-mU^\mu\nabla_\mu U_\nu+qF_{\nu\sigma}U^\sigma,\quad \theta=(mU_\nu+qA_\nu)\delta x^\nu
\end{equation}
with $E_\nu=0$ corresponding to the equation of motion for the massive charged particle in the background electromagnetic and gravitational fields.

On the other hand, the energy momentum tensor and electric current of such a point particle are defined by the response of its action to the background fields as follows
\begin{equation}
    T^{\mu\nu}\equiv\frac{2\delta S_p}{\sqrt{-g}\delta g_{\mu\nu}}=m\int d\tau U^\mu U^\nu\frac{\delta^4(x-x(\tau))}{\sqrt{-g}}, \quad j^\mu=\frac{\delta S_p}{\sqrt{-g}\delta A_\mu}=q\int d\tau U^\mu\frac{\delta^4(x-x(\tau))}{\sqrt{-g}},
\end{equation}
whereby the point particle acquires a field theory description and can be viewed as the limiting case of general matter discussed in the previous section.
By a straightforward calculation, one can show that
\begin{equation}
    \nabla_\mu j^\mu=\frac{q}{\sqrt{-g}}\int d\tau U^\mu\partial_\mu\delta^4(x-x(\tau))=-\frac{q}{\sqrt{-g}}\int d\delta^4(x-x(\tau))=0,
\end{equation}
which means that the charge is conserved as it should be the case.
The diffeomorphism invariance of the action further gives rise to
\begin{eqnarray}
    0&=&\delta_\xi S_p=\int d^4x\sqrt{-g}(\frac{1}{2}T^{\mu\nu}\mathcal{L}_\xi g_{\mu\nu}+j^\mu\mathcal{L}_\xi A_\mu)-\int d\tau E_\nu\xi^\nu-\int d[(mU_\nu+qA_\nu)\xi^\nu]\nonumber\\
    &=&\int d^4x \sqrt{-g}[\nabla_\mu(T^{\mu\nu}\xi_\nu+j^\mu A^\nu\xi_\nu)+(-\nabla_\mu T^{\mu\nu}+F^{\nu\mu}j_\mu)\xi_\nu
    -\int d\tau E_\nu\xi^\nu-\int d[(mU_\nu+qA_\nu)\xi^\nu].
\end{eqnarray}
Note that the first two terms are the field theory description of the point particle while the last two terms are the worldline description of the point particle. So the above diffeomorphism invariance relates these two distinct descriptions to each other.
In addition, it is noteworthy that the above diffeomorphism invariance is actually valid separately for each term in Eq. (\ref{pa}), so the arbitrariness of $\xi$ implies that
\begin{equation}
    \nabla_\mu T^{\mu\nu}=mU^\mu\nabla_\mu U^\nu \frac{\delta^4(x-x(\tau))}{\sqrt{-g}},
\end{equation}
which can also be obtained by a straightforward calculation. By the equation of motion $E_\nu=0$, we end up with
\begin{equation}
    \nabla_\mu T^{\mu\nu}=F^{\nu\mu}j_\mu.
\end{equation}
Last but not least, the arbitrariness of $\xi$ also implies the following identity
\begin{equation}\label{dual}
    \int_\Sigma \epsilon_{abcd}(T^{ae}+j^aA^e)\xi_e=(mU_a+qA_a)\xi^a
\end{equation}
with the right handed side evaluated at the intersection point of the worldline of the point particle with $\Sigma$. With the presence of our point particle as the perturbation onto the background fields and $\xi$ as a Killing field, Eq. (\ref{final}) together with Eq. (\ref{dual}) gives rise to
\begin{equation}
    \delta H_\xi=(mU_a+qA_a)\xi^a.
\end{equation}
In particular, when such a particle enters a black hole, then the right handed side can be interpreted as the change of black hole mass and angular moment for $\xi=\frac{\partial}{\partial t}$ and $\xi=\frac{\partial}{\partial\phi}$, respectively. Such an interpretation has been extensively exploited without derivation since long ago, especially in the test of weak cosmic censorship by the gedanken experiments starting from the seminal paper \cite{gedanken}. We have filled in this gap just now.

\section{Conclusion}
Not only is it of academic interest to sharpen the relationship between the bulk and boundary conserved quantities, but also of practical importance in gravitational physics. By working with Wald formalism, we have shown that the identification between the bulk and boundary conserved quantities, previously established in \cite{GW} for the asymptotically flat spacetimes, also applies to the asymptotically AdS spacetimes.
In addition, the perturbed matter is set to be as general as possible. In particular, there is no specific symmetry requirement imposed on the matter distribution. With this in mind, we further obtain the specific bulk boundary identification for the point particle by viewing it as the limiting case of general matter.

Although we work out the aforementioned identification explicitly only for the four dimensional spacetimes governed by the Einstein-Maxwell theory, the whole setup is obviously amenable to generalization to more generic cases, such as higher or lower spacetimes and higher derivative theories. In particular, it is interesting to obtain a similar identification for the test spinning point particle. We hope to address this issue somewhere else in the future.

We conclude our paper by pointing out that the identification we have achieved is valid only for the first order perturbations. Such an identification generically breaks down when the second order perturbations are taken into account, not to mention non-perturbatively.
\begin{acknowledgments}
This work is partially supported by the National Key Research and Development Program of China with Grant No. 2021YFC2203001 as well as the National Natural Science Foundation of China with Grant Nos. 12361141825, 12447101, 12475049, and 12575047. We like to thank Xiaokai He for his useful discussion on the conserved quantities defined on the null infinity for asymptotically flat spacetimes.

\end{acknowledgments}



\end{document}